\documentclass[twocolumn]{aa}
\usepackage{graphicx}
\usepackage{txfonts}
\usepackage{hyperref}
\usepackage{natbib}
\usepackage{soul}
\usepackage{xcolor}

\begin{document}

    \title{Uniturbulence and Alfv\'en Wave Solar Model in \texttt{MPI-AMRVAC}}
    \titlerunning{UAWSoM in \texttt{MPI-AMRVAC}}
\author{
  M. McMurdo\inst{1}\thanks{Corresponding author: \email{max.mcmurdo@kuleuven.be}} \and
  T. Van Doorsselaere\inst{1} \and
  N. Magyar\inst{2} \and
  L. Banovi\'c\inst{1} \and
  D. Lim\inst{3,1}
}

\institute{
  Centre for mathematical Plasma Astrophysics, Department of Mathematics, KU Leuven, Celestijnenlaan 200B Bus 2400, 3001 Leuven, Belgium
  \and
  Starion Group S.A., Rue des Etoiles 140, 6890 Libin, Belgium
  \and
  Solar-Terrestrial Centre of Excellence – SIDC, Royal Observatory of Belgium, Ringlaan -3- Av. Circulaire, 1180 Brussels, Belgium
}

   \date{Received ; accepted }
 
  \abstract
   {The coronal heating problem and the generation of the solar wind remain fundamental challenges in solar physics. While AWSoM-type models (Alfv\'en Wave Solar Model) have proven highly successful in reproducing the large-scale structure of the solar corona, they inherently neglect contributions from additional wave modes that arise when the effects of transverse structuring are fully incorporated into the magnetohydrodynamic (MHD) equations.}
   {In this paper, we compare the roles of kink wave- and Alfv\'en wave-driven heating in sustaining a region of the solar atmosphere, using newly developed physics and radiative cooling modules within \texttt{MPI-AMRVAC}.}
   {We extend the existing MHD physics module in \texttt{MPI-AMRVAC} by incorporating additional Alfv\'en and kink wave energy contributions to the MHD equations. We examine their roles in heating the solar atmosphere and driving the solar wind. To validate our approach, we compare numerical results from Python-based simulations with those obtained using the UAWSoM module in \texttt{MPI-AMRVAC}. Furthermore, we assess the heating efficiency of kink waves relative to that of pure Alfv\'en waves through two parameter studies: (1) exploring how different Alfv\'en wave reflection rates impact the simulated atmosphere, and (2) varying the relative magnitudes of Alfv\'en and kink wave energy injections. Finally, we present results from a larger-scale domain, sustained entirely by kink wave-driven heating.}
   {Our results show that kink wave-driven (UAWSoM) models can sustain a stable atmosphere without requiring any artificial background heating terms, unlike traditional Alfv\'en-only models. We attribute this to the increased heating rate associated with kink waves compared with Alfv\'en waves, given the same energy injection.}
   {Kink waves can sustain a model plasma with temperature and density values representative of coronal conditions without resorting to ad hoc heating terms.}

\maketitle

\section{Introduction}\label{sec:Introduction}
\nolinenumbers
The MHD equations offer a theoretical framework through which the dynamic behavior of magnetised plasma can be explored, revealing a rich spectrum of wave solutions as a response to perturbations of various plasma quantities. MHD waves have long been thought to carry energy to higher altitudes of the solar atmosphere \citep{Alfven1947} where various dissipative mechanisms are thought to convert magnetic energy into thermal energy, subsequently heating the surrounding plasma.

Perhaps the simplest solution among the wave modes supported by the MHD equations is the pure Alfv\'en wave \citep{Alfven1942}. The Alfv\'en wave solution, as is the case for all wave solutions, is not inherently straightforward due to a single perturbation generally giving rise to both forward and backward (with respect to the magnetic field) propagating wave components. This complicates the analysis of wave dynamics, particularly in nonlinear regimes where counterpropagating waves can interact nonlinearly, leading to the development of turbulence, ultimately leading to the dissipation of wave energy \citep[see, e.g.,][]{Matthaeus1999, Verdini2007, Chandran2009, Chandran2011}. To address this, the Els\"{a}sser variables \citep{Elsasser1950} have proven especially useful, enabling the decomposition of the total perturbation into two distinct fields that can be used to separately track the propagation of wave energy along and against the magnetic field direction, as well as the nonlinear interaction between counterpropagating fields.

The so-called Alfv\'en Wave Solar Model (AWSoM) proposed by \cite{Holst2014} adopted this wave frame variable approach, modeling the evolution and dissipation of Alfv\'en waves as an additional contribution to the MHD equations. These models are widely employed for space weather forecasting and have been shown to reproduce large-scale properties of the solar wind, e.g., \cite{Riley2019}. AWSoM models possess heating terms that contain both Els\"{a}sser components, which requires Alfv\'en wave reflection to generate any non-adiabatic wave heating. Alfv\'en waves can be excited in the photosphere by convective motions and propagate along the magnetic field lines, whereby stratification can cause partial wave reflection, leading to a source of counterpropagating wave energy. Numerical models often incorporate this by including an analytical reflection term in the Alfv\'en wave energy evolution equations. This reflection rate is usually defined in terms of the Alfv\'en speed gradient \citep{Holst2014, Downs2016} and dictates the proportion of Alfv\'en wave energy available for contribution to heating the plasma. Simple linear models can capture wave reflection due to the presence of longitudinal gradients in the wave speed, however, these models only show significant reflection if the length scale of the inhomogeneity is comparable to the wavelength \citep{Soler2025}. That being said, \cite{Morton2015} presented observational evidence obtained from the Coronal Multi-channel Polarimeter (CoMP) \citep{Tomczyk2007} of the existence of counterpropagating Alfv\'enic waves with around 20-80\% of the energy of outward propagating wave energy, depending on the wave frequency. Many of these numerical models, however, do not rely solely on Alfv\'en wave dissipation. They often require anomalous heating functions to produce stable atmospheres, e.g., see \cite{Mikic2018} who found insufficiently heated regions near magnetic null points and open field lines connected to weak field regions. To circumvent this, they employed two spherically symmetric heating functions to provide extra heating in the low corona. Compressive modes such as acoustic shock waves are thought to play a major role in heating the chromosphere and low corona \citep{Suzuki2005, Suzuki2006, Cranmer2007}, with Alfv\'en wave heating playing the role in heating the upper corona. These background heating terms could be thought to replicate the role of these compressive phenomena without introducing other wave modes into the domain to allow for the parameterisation of wave energy in such AWSoM(-like) models. The ability of acoustic waves to heat the chromosphere has received criticism, however, as \cite{Beck2009, Beck2012} found the wave energy to be insufficient to maintain the temperature of the atmosphere. This naturally leads to one fundamental unanswered question: what physical processes are missing in these models?

Traditionally, turbulent dissipation in magnetised plasmas has long been associated with the nonlinear interaction of counterpropagating pure Alfv\'en waves. However, the solar atmosphere is highly inhomogeneous and transverse structures have long been observed to extend far into the corona \citep[see, e.g., ][]{Raymond2014, DeForest2018, Stenborg2021}. This complexity gives rise to a broader spectrum of wave modes supported by the MHD equations \citep{Roberts1981}. \cite{Goossens2019} showed that generally, in inhomogeneous plasmas any perturbation across Alfv\'en speed isosurfaces has mixed properties. Specifically related to the presence of transverse inhomogeneity, \cite{Magyar2019} demonstrated that surface Alfv\'en waves possessed both Els\"{a}sser components and demonstrated that transverse structuring leads to a nonlinear self-deformation process known as uniturbulence, enabling wave-driven heating without requiring reflected or externally driven counterpropagating waves. \cite{Evans2012} incorporated such surface Alfv\'en waves into global MHD models and successfully reproduced key observational signatures without invoking an ad hoc heating term, highlighting the critical role of transverse structuring in models of turbulent wave heating. Extending this concept, \cite{Doorsselaere2024} developed the $Q$-variable framework, a generalisation of the Els\"{a}sser variables that allows for the decomposition of kink waves into forward- and backward-propagating components, and the authors showed that kink waves undergo a uniturbulent cascade of wave energy without the need for reflection. Much like the Els\"{a}sser formulation for Alfv\'en waves, this approach enables us to analyze turbulent wave dissipation and energy transport in transversally structured plasmas representative of the corona.

The aim of this paper is to advance upon the work conducted by \cite{Doorsselaere2025}, who numerically investigated the turbulent dissipation of kink waves in a stratified coronal plasma using a newly developed Python code, and this implementation serves as a reference implementation for the UAWSoM concept, and has been made available on Gitlab \footnote{\url{https://gitlab.kuleuven.be/plasma-astrophysics/research/tom_s-projects/uawsom.git}}. In order to build upon this work, we introduce a new physics module called the Uniturbulence and Alfv\'en wave Solar Model (UAWSoM) into the \texttt{MPI-AMRVAC} framework \citep{Xia2018, Keppens2023} to model the nonlinear turbulent dissipation of Alfv\'en and kink wave energy and the impact this has on a solar atmospheric plasma. We have made a separate repository for UAWSoM in \texttt{MPI-AMRVAC} available on Github \footnote{\url{https://github.com/maxmcmurdo/UAWSOM}}. 

The structure of this investigation is as follows, in Section \ref{sec:2}, we re-introduce the mathematical formalism introduced by \cite{Doorsselaere2024} and set out the equations implemented in this new physics module, in Section \ref{sec:3} we discuss a comparison case between the Python implementation and our new \texttt{MPI-AMRVAC} implementation and present a fully self-consistent equilibrium obtained through kink wave heating without the requirement of an ad hoc heating function. In Section \ref{sec:4} and \ref{sec:5} we conduct parameter studies in which we vary the reflection rate of Alfv\'en waves and the ratio between the Alfv\'en and kink wave energy injection, respectively. In Section \ref{sec:LargeDom1} we present simulations extending to 3 $R_{\odot}$, Section \ref{sec:Notes} discusses the potential impact of two physical effects—kink wave reflection and resonant absorption and in Section \ref{sec:Conclusions} we conclude our results and propose plans for future directions of research.

\section{Model description}\label{sec:2}

The macroscopic behavior of a plasma can be described in terms of the MHD equations. Over the top of these equations, we consider the additional effects of wave energy on the energy and momentum of the system. To model the evolution of Alfv\'en and kink wave energy in the solar atmosphere we use both the Els\"{a}sser variables given by $\vec{Z}^\pm = \vec{V} \pm \vec{v_A}$ and $Q$-variables given by $\vec{Q}^\pm = \vec{V} \pm \alpha \vec{B}$, where $\vec{V}$ represents the velocity of the fluid, $\vec{B}$ represents the magnetic field, $\vec{v_A}$ the Alfv\'en speed and $\alpha$ a generic placeholder for the speed of other waves, in this case taken to be the kink speed. Introducing these new variables allows for the formulation of the following equations

\begin{equation}
\frac{\partial \rho_0}{\partial t} + \nabla \cdot (\rho_0 \vec{V}) = 0,
\label{eq:continuity}
\end{equation}

\begin{equation}
\frac{\partial \vec{B}_0}{\partial t} - \nabla \times (\vec{V} \times \vec{B}_0) = 0,
\end{equation}

\begin{align}
    \frac{\partial (\rho_0 \vec{V})}{\partial t}
+ \nabla \cdot \left( \rho_0 \vec{V} \vec{V}
- \frac{1}{\mu} \vec{B}_0 \vec{B}_0 \right)
+ \nonumber \\ \nabla \left( p + \frac{B_0^2}{2\mu}
+ P_\mathrm{A} + P_\mathrm{k} \right)
= -\rho_0 \vec{g}(\vec{z}),
\end{align}

\begin{align}
& \frac{\partial}{\partial t} \left(
\rho_0 \frac{V^2}{2}
+ \frac{p}{\gamma - 1}
+ \frac{B_0^2}{2\mu}
+ \sum W_\mathrm{A,k}^\pm
\right)
+ \nonumber \\ & \nabla \cdot \Bigg(
\left[
\rho_0 \frac{V^2}{2}
+ \frac{p}{\gamma - 1}
+ \frac{B_0^2}{2\mu}
\right] \vec{V}
- \vec{B}_0 \frac{\vec{V} \cdot \vec{B}_0}{\mu}
\Bigg)
+ \nonumber \\ & \nabla \cdot \left(
\vec{Z}_0^- W_\mathrm{A}^+
+ \vec{Z}_0^+ W_\mathrm{A}^-
+ \vec{Q}_0^- W_\mathrm{k}^+
+ \vec{Q}_0^+ W_\mathrm{k}^-
\right)
+ \nabla \cdot \left(
[P_\mathrm{A} + P_\mathrm{k}] \vec{V}
\right) \nonumber \\ &
= - \mathcal{L}
- \rho_0 \vec{V} \cdot \vec{g}(\vec{z})
+ \frac{\zeta - 1}{\zeta + 1}
P_\mathrm{k} \nabla \cdot \vec{V},
\label{eq:systemenergy}
\end{align}

\begin{equation}
\frac{\partial W_\mathrm{A}^\pm}{\partial t}
+ \nabla \cdot (\vec{Z}_0^\mp W_\mathrm{A}^\pm)
+ \frac{W_\mathrm{A}^\pm}{2} \nabla \cdot \vec{V}
= - \Gamma^\mp W_\mathrm{A}^\pm \mp \mathcal{R}_\mathrm{A},
\label{eq:alfven}
\end{equation}

\begin{equation}
\frac{\partial W_\mathrm{k}^\pm}{\partial t}
+ \nabla \cdot (\vec{Q}_0^\mp W_\mathrm{k}^\pm)
+ \frac{W_\mathrm{k}^\pm}{2} \nabla \cdot \vec{V}
= - \frac{1}{L_{\perp,\mathrm{VD}}}
\frac{1}{\sqrt{\rho_\mathrm{e}}}
(W_\mathrm{k}^\pm)^{3/2}\mp \mathcal{R}_\mathrm{k},
\label{eq:kink}
\end{equation}
where $\mu$ is the permeability constant, $p$ the hydrodynamic pressure, $P_{A,k}$ are the additional pressure contributions from Alfv\'en and kink waves, respectively. These expressions are defined by \cite{Holst2014, Doorsselaere2025} to be

\begin{equation}
P_\mathrm{A}=\frac{W^+_\mathrm{A}+W^-_\mathrm{A}}{2}, \quad \mbox{and} \quad
P_\mathrm{k}=\frac{W^+_\mathrm{k}+W^-_\mathrm{k}}{2\mu\rho_0\alpha^2}.
\label{eq:wave_pressure}
\end{equation}
Continuing the adopted notation, $\vec{g}$ represents gravity, $\gamma$ is the polytropic index taken to be $5/3$ (corresponding to an ideal monatomic gas with the pressure and density related via $p \propto \rho^\gamma$), $W_\mathrm{A,k}^\pm$ represents the Alfv\'en and kink wave energies for inward and outward propagating waves, $\vec{Z}_0$ and $\vec{Q}_0$ represent the equilibrium Els\"{a}sser and $Q$-fields, corresponding to the Alfv\'en and kink speeds, respectively (with $\vec{B}_0$ the equilibrium magnetic field), $\mathcal{L}$ the radiative losses, $\zeta$ the density contrast of the plasma between the inside and outside of the coronal plume (prescribed later), $\Gamma^\pm$ are the heating rates for Alfv\'en waves (defined later), the correlation length scale for kink waves are given by $L_{\perp,\mathrm{VD}}$ and $\mathcal{R}_\mathrm{A,k}$ represents the Alfv\'en and kink wave energy reflection term. In this initial implementation, we set the reflection terms ${\cal R}_\mathrm{k}$ in the kink wave energy equation (Equation \ref{eq:kink}) to zero. Their correct values are not well determined \citep{Pelouze2023} and will be investigated in future work. Alfv\'en wave reflection is crucial to obtain direct Alfv\'en wave heating, and this will be discussed in more detail later. In Equations (\ref{eq:continuity}-\ref{eq:wave_pressure}), the density, $\rho_0$ is the weighted average between the external ($\rho_\mathrm{e}$) and internal density ($\rho_\mathrm{i}$) of the assumed transverse structure (e.g., a plume). In order to perform our investigation of kink wave propagation, we collapse the dimensions of what is inherently a 3D structure, down to a single dimension. To do this, we must calculate the average density over the cross-section, weighted by the density of the structure and its surroundings. When averaging the density over a cross-section, we have

\begin{align}
& \frac{f}{\pi R^2} \int d\phi \int_0^{\frac{R}{\sqrt{f}}} \rho rdr = \frac{f}{\pi R^2} \left( \pi R^2 \rho_\mathrm{i} + \pi \left(
\frac{R^2}{f}-R^2\right) \rho_\mathrm{e}\right)= \nonumber \\ & f\rho_\mathrm{i} + (1-f)\rho_\mathrm{e}=\rho_\mathrm{e}+f(\rho_\mathrm{i}-\rho_\mathrm{e}) = \rho_0,
\end{align}
where $R$ represents the radius of the fine scale structure within a coronal plume, often referred to as micro-plumes \citep{Wilhelm2011} or plumelets \citep{Uritsky2021}, which expands according to the magnetic field with height and $f$ is the filling factor taken to be 0.1 throughout the present investigation \citep[as introduced by ][]{TVD2014}. This value quantifies the area occupied by the plumelets as a fraction of the total volume and is assumed to be constant. We note that investigations by \cite{Morton2023} show that the corona is structured down to the smallest observable scales with a power law dependence on scale, which could result in the filling factor being larger than 0.1. Altering the filling factor in our simulations leads to changes in the averaged density, kink wave correlation length, wave speed and hence wave flux, leading to altered heating rates and, of course, the radiative losses. 

The Alfv\'en heating terms are taken from \cite{Holst2014} and are given by 

\begin{equation}
    \Gamma^\mp = \frac{2}{L_{\perp,\mathrm{AW}}}\sqrt{\frac{W_\mathrm{A}^\mp}{\rho_0}},
\end{equation}
where we \citep[as][]{Holst2014} take $L_{\perp,\mathrm{AW}} = 1.5\times 10^5 \sqrt{\frac{T}{B}}$, in which $T$ is the temperature of the plasma, given in MK, $B$ the total magnetic field and the factor at the beginning of the expression is given such that the total expression is given in meters. \cite{Downs2016} and \cite{Reville2020} offer an alternative definition, wholly dependent on the magnetic field, and these authors used a different constant, which may be varied to produce stronger or weaker damping that could potentially be used to capture a frequency dependence on the wave energy injection. \cite{Doorsselaere2024} introduced a correlation length for kink waves that is dependent on the physical structure of the plasma, and is shown to vary with the density contrast, radius of the plume and filling factor and is given by

\begin{equation}
    L_{\perp,\mathrm{VD}}(R,\zeta,f) = \left(\frac{\sqrt{2}(\zeta - 1)}{2R\sqrt{5f\pi}} \frac{1-f^{5/2}}{(\zeta+1-f)^{3/2}}\right)^{-1}.
\end{equation}
The dependence on density contrast is physically expected, since larger density contrasts naturally lead to the onset of smaller spatial scales \citep{Soler2015}. This effect is similar to the enhanced dissipation of Alfv\'en wave damping due to phase mixing \cite[e.g., see,][]{HP1983, McMurdo2023, McMurdo2025}.

Special care must be taken when calculating the radiative losses of the plasma. Since we consider an averaging process to calculate a single density, we must carefully consider the relative contributions to the radiative losses from the internal and external regions of our plume. So, in addition to developing the UAWSoM module, we modified the existing \texttt{MPI-AMRVAC} radiative cooling module to better capture the locality of radiative losses while still employing an averaging process for the assumed density structure. This approach allows us to solve the one-dimensional evolution of the atmosphere more accurately.

The average of the squared density over both the interior and the exterior of the plume, $\langle\rho_0^2\rangle$, is an important quantity to be defined as this dictates the averaged radiative losses of the plasma associated with the density contrast across the plume. To derive this quantity, we begin with the expression for the total averaged density, $\rho_0$, given by
\begin{align*}
\rho_0 &= (1 - f)\rho_\mathrm{e} + f\rho_\mathrm{i} \\
&= \rho_\mathrm{e} \left[(1 - f) + f\zeta\right] \\
&= \rho_\mathrm{e} (1 - f + f\zeta).
\end{align*}
Now, we can obtain expressions for the square of the internal and external densities in terms of the averaged density
\begin{align*}
\rho_\mathrm{e}^2 &= \frac{\rho_0^2}{(1 - f + f\zeta)^{2}} \\
\rho_\mathrm{i}^2 &= \zeta^2 \rho_\mathrm{e}^2.
\end{align*}
We can then take the sum of these values weighted by the filling factor such that
\begin{align*}
    \langle \rho_0^2 \rangle &= (1-f) \rho_\mathrm{e}^2 + f\rho_\mathrm{i}^2 \\
    &= \rho_0^2\frac{(1-f) + \zeta^2 f}{(1 - f + f\zeta)^{2}} \\ &= \rho_0^2 \left[1 + \frac{f(1 - f)(\zeta - 1)^2}{(1 + f\zeta - f)^2} \right].
\end{align*}
In summary, the factor alongside $\rho_0^2$ accounts for the difference between the average of the squared density and the square of the averaged density, rooted in the density contrast between the interior and the exterior of a plume.

Calculation of the radiative losses is based on the implementation of \cite{Hermans2021}, and the radiative loss function is given by
\begin{equation}\label{eq:radiative_cooling}
    \mathcal{L} = \frac{\langle\rho_0^2\rangle}{(1+4A_{\text{He}})^2m_p^2}\Lambda(T).
\end{equation}
Here $A_{\text{He}}$ represents the ratio of helium to hydrogen abundance, set by default to 0.1, while $m_p$ is the proton mass. The cooling curve $\Lambda(T)$ is calculated using the data from \cite{Colgan2008}. We interpolate the tabulated cooling curve values to the temperatures required during the simulation using a cubic spline approach. 

In addition to the terms already present in the energy equation (Equation \ref{eq:systemenergy}), we include a thermal conduction contribution on the right-hand side (RHS):
\begin{equation}
    \mathrm{Thermal\ conduction} = - \nabla \cdot (\kappa \nabla T),
\end{equation}
where $\kappa = \kappa_0 T^{5/2}$, with $\kappa_0 = 8 \times 10^{-7}\ \mathrm{erg\ cm^{-1}\ s^{-1}\ K^{-7/2}}$. We neglect the effects of perpendicular thermal conduction, as it is typically several orders of magnitude lower than the parallel component due to the strong guiding influence of the magnetic field in the corona \citep{Braginskii1965}. This anisotropy allows thermal energy to be transported efficiently along field lines while effectively inhibiting cross-field transport. 

\section{Comparison between \cite{Doorsselaere2025} and \texttt{MPI-AMRVAC} simulations}\label{sec:3}

Recently, \cite{Doorsselaere2025} implemented the $Q$-variable formulation of the MHD equations in a Python code and found that kink waves, were able to some extent, counterbalance the radiative losses, delaying the onset of catastrophic cooling. In this section, we present some corroborating results between the two implementations of the same equations, one obtained from the Python implementation \citep{Doorsselaere2025}, based on the {\texttt{fipy}} package\footnote{\url{https://www.ctcms.nist.gov/fipy/}}\citep{Guyer2009}, and the \texttt{MPI-AMRVAC} implementation. The boundary conditions are chosen to be the same throughout the two implementations of UAWSoM, albeit in different ways, more details to follow. We also wish to draw the readers' attention to some slight modifications made to the Python implementation in \cite{Doorsselaere2025}. The boundary conditions for density differ in that now, we extrapolate the density, assuming a hydrostatic equilibrium at both boundaries, rather than considering an open boundary. In addition to this, we opted for a new cooling curve given by \cite{Colgan2008} for improved numerical stability, and we found this to play no major role in either the Python or \texttt{MPI-AMRVAC} result.

\subsection{Boundary conditions}

In the most general form employed throughout the present investigation, the boundary conditions at the bottom boundary are given by:
\begin{align}
    \frac{d\rho_0}{dz} &= \vec{g}(z)\frac{\rho_0^2}{p},\\
    \frac{dp}{dz} &= \rho_0 \vec{g}(z),\\
    \frac{dV}{dz} & = 0, \\
    W_\mathrm{k}^- & = W_\mathrm{k0}, \\
    W_\mathrm{A}^- & = W_\mathrm{A0}, \\
    \frac{dW_\mathrm{k}^+}{dz} & = 0, \\
    \frac{dW_\mathrm{A}^+}{dz} & = 0,
\end{align}
where $W_\mathrm{A0,k0}$ are the values of Alfv\'en and kink wave energy injections at the base of the domain, respectively. Gravity is assumed here to be constant (to allow for comparison between the two implementations) and is given by $\vec{g}(z) = -274\ \mathrm{m\ s^{-2}}\ \vec{1}_z$. At the top boundary, we put as boundary conditions
\begin{align}
    \frac{d\rho_0}{dz} &= \vec{g}(z)\frac{\rho_0^2}{p},\\
    \frac{dp}{dz} &= \rho_0 \vec{g}(z),\\
    \frac{dV}{dz} & = 0, \\
    \frac{dW_\mathrm{k}^-}{dz} & = 0,\\
    \frac{dW_\mathrm{A}^-}{dz} & = 0,\\
    W_\mathrm{k}^+ & = 0.
\end{align}
which expresses a continuous isothermal stratification beyond the computational domain through the extrapolation of density and pressure. The low plasma beta and high thermal conductivity along magnetic field lines in the corona help maintain a relatively stratified, quasi-static atmosphere over large spatial scales. Although the background atmosphere itself may develop to become non-isothermal on a global scale, assuming local isothermality in the boundary regions simplifies the enforcement of hydrostatic balance while maintaining compatibility with the evolving coronal structure. For the wave energy and velocity, open boundary conditions are employed to realistically mimic the largely open and dynamic nature of the coronal environment, allowing for mass to be replaced after the wave pressure drives plasma out of the domain, allowing for long-term simulations to be performed. These boundary conditions allow waves and perturbations to leave the computational domain without artificial reflections, which could otherwise introduce nonphysical interference patterns and energy buildup. Together, these assumptions ensure a stable and physically consistent background over which we model the propagation of Alfv\'en and kink wave energy and their associated dissipation processes in the corona.

\subsection{Initial conditions}

For the cases discussed here, we utilise a pre-implemented plasma configuration with initial profiles given by:
\begin{align}
    \rho_0(z) &= \rho_0(0) \exp\left(-\frac{z}{H}\right), \label{eq:background_rho}\\
    V(z) &= 0, \label{eq:V_of_height}\\
    B(z) &= B_0 \frac{R_\sun^2}{\left(z + R_\sun\right)^2}, \label{eq:background_B}\\
    R(z) &= R_0 \sqrt{\frac{B_0}{B(z)}}, \label{eq:background_R}\\
    \zeta(z) &= (\zeta_0 - 1)\exp\left(-\frac{z}{5 R_\sun}\right) + 1, \label{eq:background_zeta}
\end{align}
where we take $\rho_0(0)$ corresponding to a number density of $10^9\ \mathrm{cm^{-3}}$, $H = 50\ \mathrm{Mm}$, $B_0 = 20\ \mathrm{G}$ (the base magnetic field intensity), $\zeta_0 = 5$, $R_0 = 1\ \mathrm{Mm}$ and an initial temperature is set uniformly to $T_0 = 1\ \mathrm{MK}$ throughout the domain. These parameters, while not necessarily representative of a coronal plume, were chosen to allow for the direct comparison with results presented in \cite{Doorsselaere2025}. We note that decreasing the density contrast will reduce the total radiative losses, resulting in lower energy injection requirements for a given temperature. It will also reduce the kink wave heating term, albeit minimally (for example, for $\zeta_0 = 2$, and all other parameters remaining the same, the heating rate achieves 70\% of the current value), while reducing the magnetic field will reduce the flux of energy. For our chosen parameters, we find that the kink wave correlation length, $L_{\perp,\mathrm{VD}}$, ranges between $6.4 - 7.2$ Mm matching closely with those presented in \cite{Sharma2023} who found a distribution that sharply peaked around $7.6 - 9.3$ Mm (with values ranging between $2 - 20$ Mm). We find that at lower heights, the correlation length of Alfv\'en and kink waves closely match, which is subject to change for larger heights when the assumed density contrast approaches unity. The choices of magnetic field intensity and density contrast have important influences on the wave flux, radiative losses and correlation length of kink waves, and while a full parameter study is not the focus of the present investigation, stable (quasi-equilibrium) atmospheres are generally achievable for a large range of values by varying the injected wave energy. Drawing on observational results, we estimated the kink wave energy injection at the base of the domain to be $W_\mathrm{k0} = 5.3 \times 10^{-3}\ \mathrm{J\ m^{-3}}$ by adopting Equation (15) from \citet{TVD2014}, as given below (in a new form relative to the present investigation):
\begin{equation}\label{eq:energydensity}
W_\mathrm{k} = \frac{1}{2}f\rho_0 |\delta Q|^2,
\end{equation}
where $|\delta Q|$ is the velocity amplitude of the oscillation, i.e., the magnitude of the $Q$-variable perturbation. \citet{Weberg2020} detected, on average, around 600 transverse waves in coronal plumes at altitudes between 5 and 35~Mm, and reported an average velocity amplitude of approximately 10 km~s$^{-1}$ at about 5~Mm. Given the substantial evidence for the ubiquity of transverse waves in open field regions \citep{Morton2015, Morton2016}, we estimate their possible total energy density, representing the kink wave energy input at the base of our domain, by assuming that the detected waves are independent and share the same velocity amplitude. 

\begin{figure}
    \centering
    \includegraphics[width=0.49\textwidth]{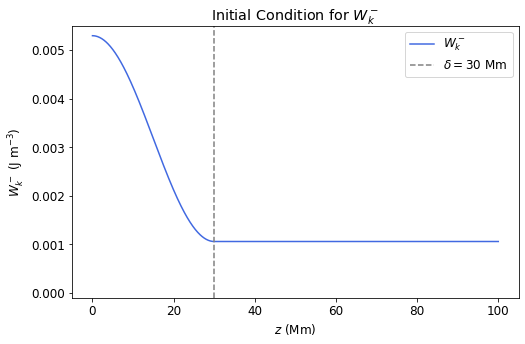}
    \caption{The initial profile of outward propagating kink wave energy is shown as a function of height over the $100$ Mm domain. Beyond $30$ Mm, the initial wave energy is assumed to be constant.}
    \label{fig:Wkm_IC}
\end{figure}

The initial conditions for the wave energy densities $W^\pm_\mathrm{A,k}$ are given by:
\begin{align}
    W^-_\mathrm{k} &=
    \begin{cases}
        0.4 W_\mathrm{k0}\left(\cos\left(\dfrac{\pi z}{\delta}\right) + 1.5\right), & \text{for } z < \delta, \\
        0.2 W_\mathrm{k0}, & \text{otherwise},
    \end{cases} \\
    W^+_\mathrm{k} &= W^\pm_\mathrm{A} = 0,
\end{align}
where $\delta = 30\ \mathrm{Mm}$. The smooth profile for $W^-_\mathrm{k}$ is chosen to avoid numerical artifacts that could result from abrupt discontinuities between zero and nonzero values. The subsequent injection of wave energy into the domain is solely at the bottom boundary. This initial condition is shown in Figure \ref{fig:Wkm_IC}. 

\subsection{Simulation comparison}

\begin{figure*}[htp]
    \centering
    \begin{minipage}[t]{0.49\textwidth}
        \centering
        \includegraphics[width=\textwidth]{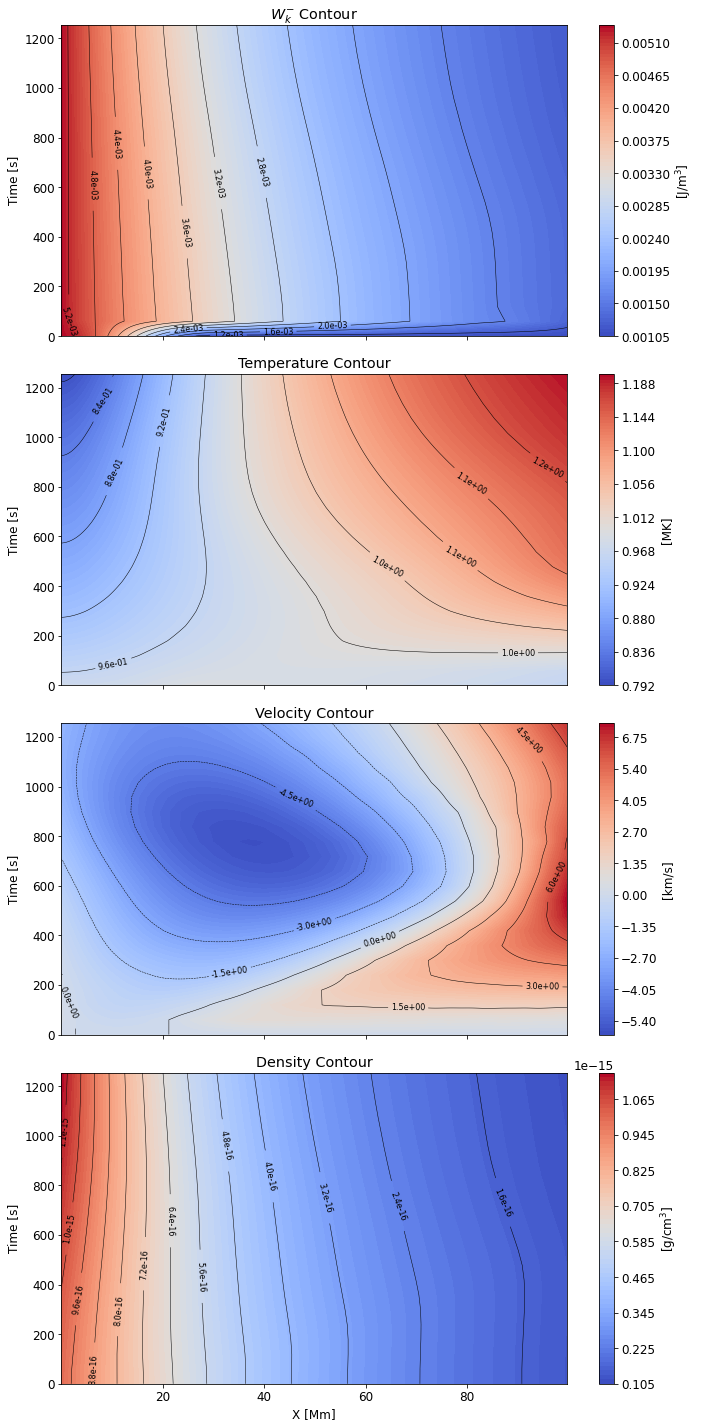}
    \end{minipage}
    \hfill
    \begin{minipage}[t]{0.49\textwidth}
        \centering
        \includegraphics[width=\textwidth]{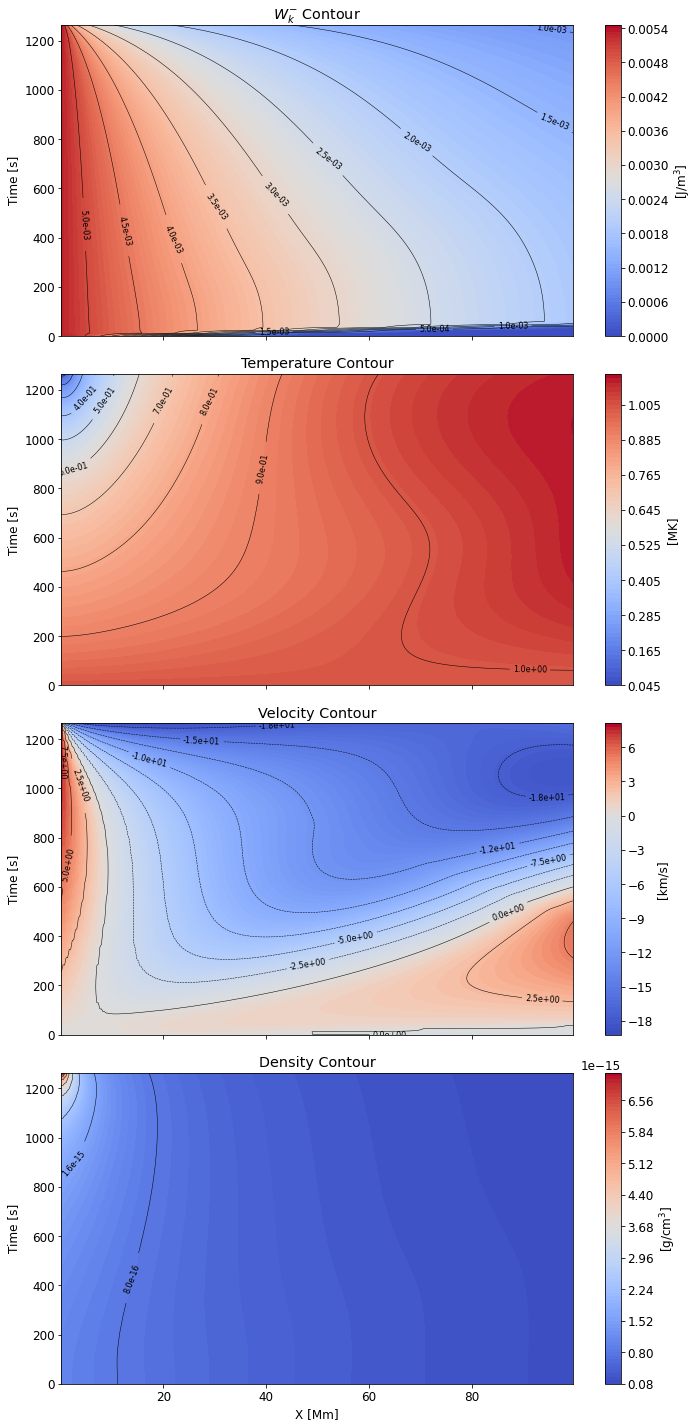}
    \end{minipage}
    \caption{From the top panel, working downwards, we present the results obtained from the new module UAWSoM in \texttt{MPI-AMRVAC} for the outward propagating kink wave energy ($\mathrm{J\ m^{-3}}$), temperature (in units MK), the velocity (in km\,s$^{-1}$) and the plasma density (in units g\,cm$^{-3}$) as functions of time (seconds) and space, $X$ (Mm). The left-hand column represents the results obtained from the \texttt{MPI-AMRVAC} implementation of UAWSoM, while the data presented in the right-hand column is obtained from the Python implementation.}
    \label{fig:Py_AMRVAC_comp}
\end{figure*}

Results obtained from both the Python and \texttt{MPI-AMRVAC} simulations of outwardly propagating kink wave energy, temperature, velocity, and density over the same time span and setup are compared in Figure \ref{fig:Py_AMRVAC_comp}. In these simulations, the Alfv\'en wave energy is set to zero to isolate the evolution of kink wave energy and its impact on the coronal plasma.

The observed differences are primarily attributed to variations in how identical boundary conditions are implemented. Specifically, \texttt{MPI-AMRVAC} uses ghost cells to extrapolate physical quantities beyond the domain, whereas the Python version lacks this capability. This means that \texttt{MPI-AMRVAC} can properly account for energy outside of the domain, resulting in stable inflows. Furthermore, \texttt{MPI-AMRVAC} has adaptive time-stepping, which helps manage steep gradients during the relaxation phase, a feature not present in the Python implementation. Differences in how thermal conduction and radiative losses are treated also contribute to the discrepancies. However, after extensive testing (not shown), including runs without radiative losses and/or thermal conduction, we conclude that boundary condition treatment is the dominant factor. The goal here is not to assess the Python code, but to validate the results of \cite{Doorsselaere2025}, prior to further development of the current \texttt{MPI-AMRVAC} module.

Due to this variation in boundary behaviour, the two simulations exhibit variation in the flow of plasma, resulting in differing levels of adiabatic heating and/or cooling. Adiabatic heating and cooling processes are likely to account for the majority of discrepancies observed between the two simulations' results. In the \texttt{MPI-AMRVAC} simulations, condensations develop in the lower half of the domain, resulting in downflows, while wave pressure is significant enough to induce upflows in the upper half of the domain. This dispersion of mass leads to additional cooling effects beyond those attributable to radiative losses. In contrast, the Python-based simulations exhibit strong upflows entering from the lower boundary, with an effect on the outcome of the simulation. These upflows are not correspondingly balanced at the upper boundary—an inconsistency that becomes significant when considering the stratified nature of the plasma. 

Despite these differences in dynamics, both simulations display broadly similar temperature profiles, suggesting the dominant heating mechanism remains kink wave heating. Plasma cooling is most pronounced near the base of the domain, where the density is greatest, while temperatures at the top boundary exceed the initial uniform temperature of 1 MK.

\subsection{Steady state UAWSoM solution}

\begin{figure}[htp]
   \centering
    \includegraphics[width=.49\textwidth]{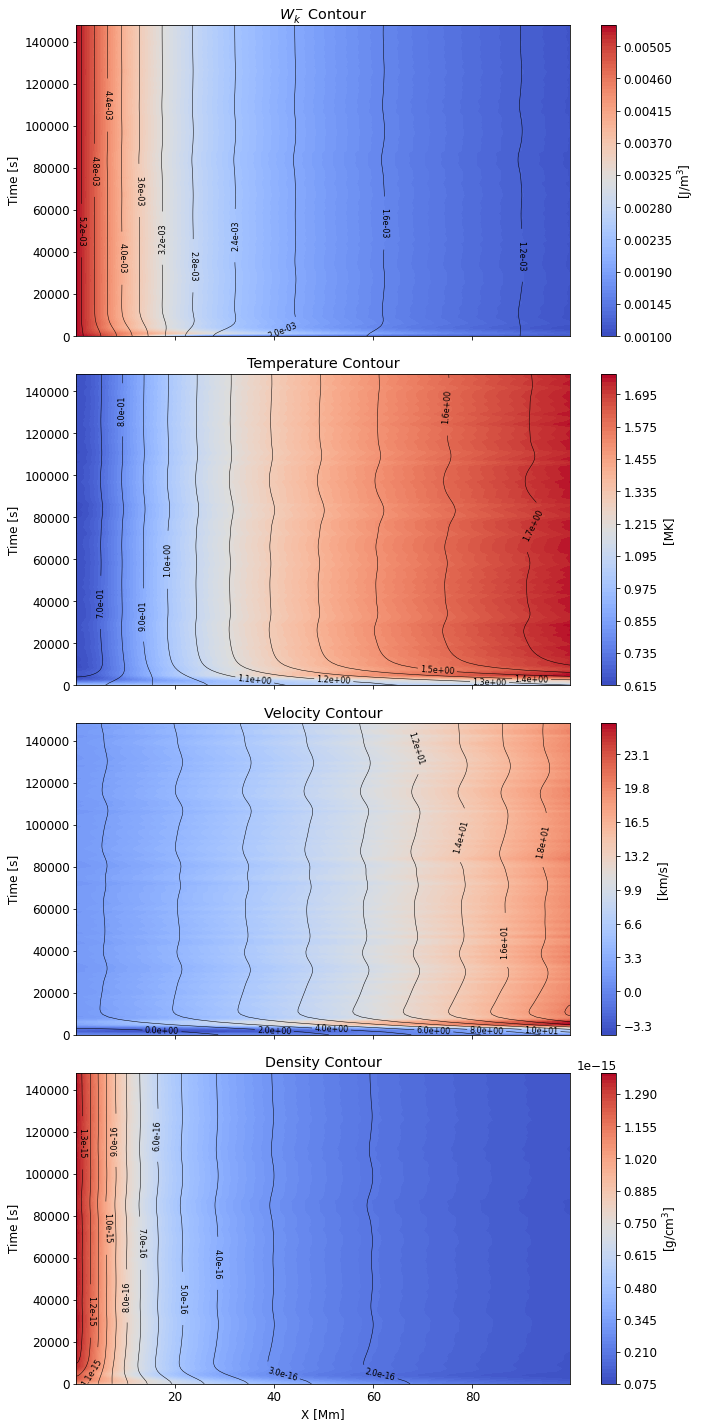}
    \caption{\texttt{MPI-AMRVAC} simulation as prescribed in Figure \ref{fig:Py_AMRVAC_comp} but now the simulation is allowed to run for much longer. We consider this to be a qausi-steady state. The quantities shown are the outward propagating kink wave energy ($\mathrm{J\ m^{-3}}$), temperature (in units MK), the velocity (in km\,s$^{-1}$) and the plasma density (in units g\,cm$^{-3}$) as functions of time (seconds) and space, $X$ (Mm).}
  \label{fig:AMRVAC_kink_only_steady}  
\end{figure}

One of the fundamental questions we answer in this study is: Can we maintain a stable model atmosphere without anomalous background heating? Below, we present such a result. A stable atmosphere without background heating that exhibits coronal temperatures that are maintained against radiative losses solely by kink wave heating. The straight contour lines in Figure \ref{fig:AMRVAC_kink_only_steady} show that the system has reached a steady state. To obtain the results in Figure \ref{fig:AMRVAC_kink_only_steady}, we simply ran the above simulation for longer. 

The solar corona is an optically thin plasma and is characterised by a typical temperature of \( T \approx 10^6\,\mathrm{K} \) and number density \( n \approx 10^9\,\mathrm{cm^{-3}} \). These conditions correspond to an internal energy density of approximately \( e \sim 2 \times 10^{-2}\,\mathrm{J\,m^{-3}} \). The associated optically thin radiative loss rate is on the order of \( Q_{\mathrm{rad}} \sim 10^{-5}\,\mathrm{W\,m^{-3}} \), yielding a characteristic radiative cooling timescale of
\[
\tau_{\mathrm{cool}} = \frac{e}{Q_{\mathrm{rad}}} \approx 2000\,\mathrm{s},
\]
which is typically the longest relevant physical timescale in the coronal plasma. Our temporally extended simulation (presented in Figure \ref{fig:AMRVAC_kink_only_steady}) is run over timescales much longer than any of these characteristic values, ensuring that the system has sufficient time to evolve dynamically and thermodynamically. This allows for a proper investigation of quasi-steady-state coronal configurations and wave heating processes.

As Figure \ref{fig:AMRVAC_kink_only_steady} demonstrates, the kink wave heating is entirely responsible for maintaining coronal plasma temperatures $\sim1$ MK. The steady outflow at the top boundary is balanced by a steady inflow (as shown in the constant density and velocity profile in time), suggesting there is negligible adiabatic heating occurring in our simulation. Satisfied with the implementations of the set of governing equations within the \texttt{MPI-AMRVAC} framework, we now present a collection of results obtained through a parameter study of the various user defined variables. 

\section{Parameter study: varied Alfv\'en wave reflection rates}\label{sec:4}

In the previously discussed AWSoM models, it is a necessity that the equations produce a source of reflected Alfv\'en waves. There is some variation in the literature on what this reflection should be. For example, \cite{Reville2020} used a constant reflection rate of $0.1$, while \cite{Holst2014} uses a more involved function involving the curl of the velocity field and the gradient of the Alfv\'en speed. \cite{Downs2016} used a simpler reflection rate dependent on the gradient of the Alfv\'en speed, which can be obtained by considering the one-dimensional case of \cite{Holst2014}, where naturally the curl of the velocity field is zero. This term is analogous to the work of \cite{Heinemann1980} who calculated the non-Wentzel–Kramers–Brillouin (WKB) reflection rate in a stratified plasma, only now \cite{Downs2016} ignored the effect of the radial acceleration of the wind on reflection. If the equations are written in terms of wave energy,  \cite{Holst2014} and \cite{Downs2016} require a `seed' of inward propagating wave energy for reflection to occur, since the terms corresponding to reflection contain both inward and outward propagating wave energy. This means that unidirectionally propagating wave energy does not self-consistently produce a source of counterpropagating wave energy, which is a result of the parameterisation of the wave energy necessary to perform global models. It is not our goal to reproduce the results previously obtained, but rather to evidence the relative capabilities of Alfv\'en and kink waves to heat a coronal plasma. For this reason, we choose a reflection rate that can produce reflected waves in a self-consistent way without the requirement of initial wave energy propagating in both directions. We take inspiration from the aforementioned literature, and the reflection term we consider in Equation (\ref{eq:alfven}) is given by

\begin{equation}
    \mathcal{R}_\mathrm{A} = \sigma\left(\frac{V + v_A}{v_A}\right)\frac{dv_A}{dz}W_{\mathrm{A}}^\pm,
    \label{eqn:Reflection_rate}
\end{equation}
where $\sigma$ is a constant that we vary between simulations to simulate stronger or weaker Alfv\'en wave energy reflection. The quantity in the brackets reflects the importance of the background flow on reflection and was derived from \cite{Downs2016}. The approach we take in which we allow $\sigma$ to be a free parameter, is similar to \cite{Matthaeus1999} and \cite{Chandran2009}, who conducted parameter studies where the reflection rate was varied between simulations. This approach aims to quantify the significance of an Alfv\'en speed gradient on the reflection of Alfv\'en wave energy. Varying $\sigma$ is motivated by assuming varied frequency compositions of the wave energy and the impact this would have on reflection \citep{Soler2025}. This places more or less emphasis on Alfv\'en wave heating as the dominant contributing factor to the overall plasma heating and can shift the distribution of heating within the domain.

\subsection{Simulation results}\label{sec:sameAWK}

\begin{figure}[htp]
   \centering
    \includegraphics[width=.45\textwidth]{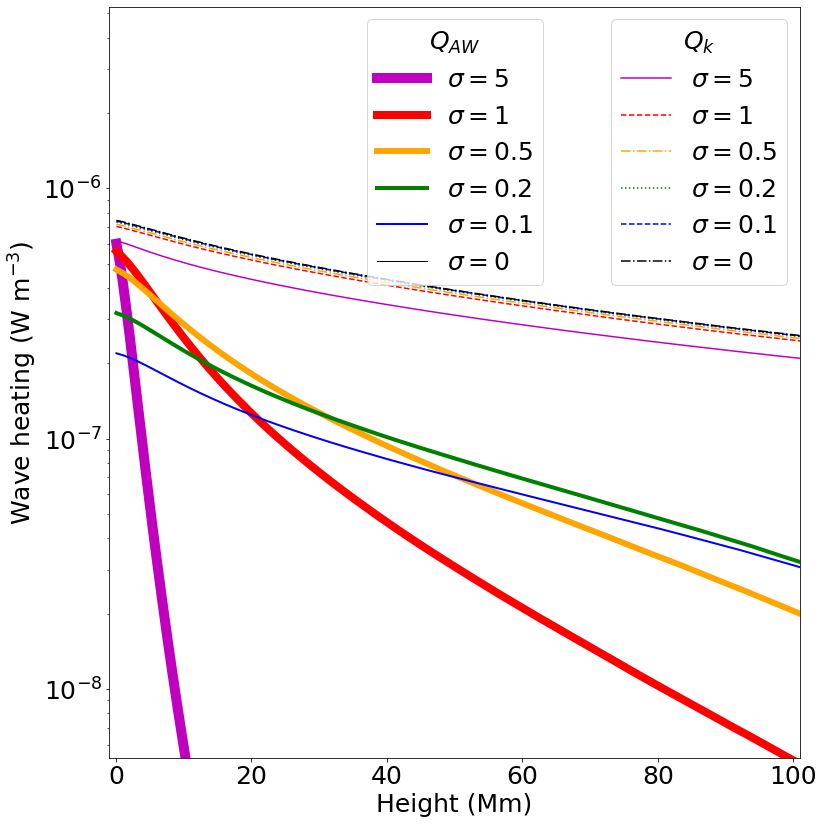}
    \caption{The heating rates (given in W\,m$^{-3}$) corresponding to Alfv\'en and kink waves given the same energy injection are shown as functions of height for the various reflection rates considered. The reflection coefficients ($\sigma$) are 5 (pink), 1 (red), 0.5 (orange), 0.2 (green), 0.1 (blue) and 0 (black) that leads to no Alfv\'en wave heating. The various thicknesses of lines refer to the Alfv\'en wave heating term, and the various line styles refer to the kink wave heating terms.}
  \label{fig:AMRVAC_varyAWK_ref_heating_sameinj}  
\end{figure}

Below, Figures \ref{fig:AMRVAC_varyAWK_ref_Tv1_sameinj} and \ref{fig:AMRVAC_varyAWK_ref_heating_sameinj} show the temperature and velocity profiles as functions of height and the Alfv\'en wave and kink wave heating contributions, respectively. The values of $\sigma$ inserted into the reflection function given by Equation (\ref{eqn:Reflection_rate}), are given by $\sigma = 0$ (black),  0.1 (blue), 0.2 (green), 0.5 (orange), 1 (red) and 5 (pink). For these simulations, we considered the energy injections of the two waves to be initially identical and exactly one half of the value employed in Section \ref{sec:3}, such that the sum of the two wave energy injections at the base of the domain achieve the value given by $W_\mathrm{k0}$. The initial energy injection is supplemented by an identical injection at the base (i.e., while the damping profiles of the two waves can vary, their base injection remains the same throughout the simulation). Given the same energy input, kink waves and Alfv\'en waves do not have the same heating rates, the kink wave heating term is proportional to the kink wave energy to the power of three halves, while the heating due to Alfv\'en waves is proportional to one of the inward or outward propagating Alfv\'en wave energies multiplied by the other signed Alfv\'en wave energy to the power of one half. Given the similarity between the correlation length scales for kink waves and Alfv\'en waves, at least in the lower atmosphere (typically of the order of a few percent of a $R_{\odot}$), we simplify the forthcoming explanation of the disparity in heating rates by neglecting the impact of differing dissipation lengths. The reason for the discrepancies then comes from the condition that Alfv\'en waves can only heat the plasma if both inward and outward propagating waves coexist at the same spatial position. Throughout the present study, we consider that inward propagating Alfv\'en wave energy can only be generated through the reflection of Alfv\'en wave energy that originated as outwardly propagating Alfv\'en wave energy. In the limit of full Alfv\'en wave reflection, all of $W_\mathrm{A}^-$ can be fed into $W_\mathrm{A}^+$, resulting in the heating rates for Alfv\'en and kink waves to match at this infinitesimally small reflective point, provided this point is the first and only location of reflection within the domain. In every case less extreme than full reflection, Alfv\'en wave heating is strictly less efficient than kink wave heating at every spatial point, thus highlighting the efficiency of uniturbulent dissipation in heating the structured solar atmosphere, especially in weakly stratified plasmas, where reflection is minimal.
 
The discrepancies in the heating rates explain why Figure \ref{fig:AMRVAC_varyAWK_ref_Tv1_sameinj} exhibits only minor variations in temperature as we vary the Alfv\'en wave energy reflection rate; the dominant heating contribution remains associated with kink waves. In Figure \ref{fig:AMRVAC_varyAWK_ref_heating_sameinj}, only in the case of exceptionally strong reflection ($\sigma = 5$, pink line) does the Alfv\'en wave heating approach the magnitude of the kink wave heating near the lower boundary. It is important to note, however, that even in this extreme scenario, Alfv\'en wave heating in the upper atmosphere remains negligible. Furthermore, the enhanced downward-propagating Alfv\'en wave energy associated with strong reflection introduces a destabilising influence on the atmosphere (not associated with a lack of wave energy, but rather an excess of localised wave pressure), ultimately leading to catastrophic cooling. This is the reason we present results at $t = 1800$ s and cannot extend the simulation beyond this time. We note that while Alfv\'en and kink wave heating do not directly interact with one another, the influence that each wave has on plasma quantities, such as density, results in the kink wave heating differing from one simulation to the other. This is shown in Figure \ref{fig:AMRVAC_varyAWK_ref_heating_sameinj}, by the variation in the kink wave heating profile, $Q_\mathrm{k}$.

\begin{figure}[htp]
   \centering
    \includegraphics[width=.45\textwidth]{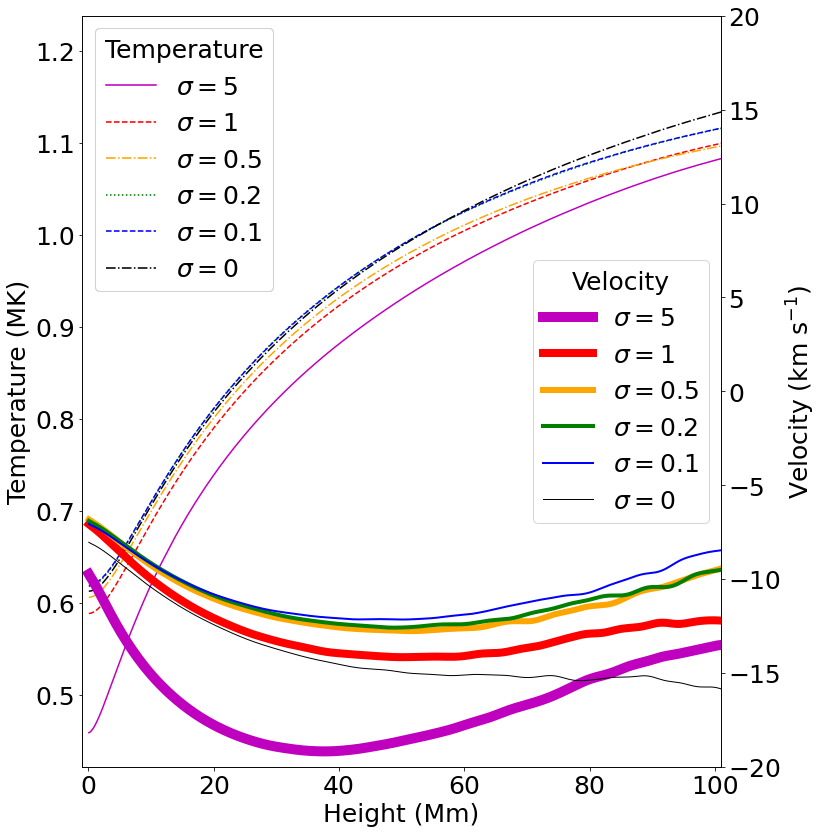}
    \caption{Temperature (in MK given by various line styles and colors) and velocity (in km\,s$^{-1}$ given by various thickness and color solid lines) are given as a function of height (Mm) plotted at time $t = 1800$ s, where the reflection coefficient ($\sigma$) corresponding to Alfv\'en wave energy reflection is varied. The reflection coefficients ($\sigma$) are 5 (pink), 1 (red), 0.5 (orange), 0.2 (green), 0.1 (blue) and 0 (black) that leads to no Alfv\'en wave heating. The various thicknesses of lines refer to the velocity, and the various line styles refer to the temperature.}
  \label{fig:AMRVAC_varyAWK_ref_Tv1_sameinj}  
\end{figure}

\subsection{Increased Alfv\'en wave injection}

Due to this inefficiency in Alfv\'en wave heating, we now choose to increase the Alfv\'en wave energy injection at the base of the domain, such that the heating rates for Alfv\'en and kink waves are comparable. To choose this value, we calculated an approximate efficiency ratio between Alfv\'en and kink waves. The heating rate corresponding to Alfv\'en waves, \( Q_{\mathrm{AW}} \propto W_\mathrm{A}^- \sqrt{W_\mathrm{A}^+} \) and since inward propagating Alfv\'en wave energy, \( W_\mathrm{A}^+ \) is generated solely within the domain through the reflection (and consequent depletion) of inward propagating Alfv\'en wave energy, \( W_\mathrm{A}^- \), it is possible to rewrite the heating rate in terms of outward propagating Alfv\'en wave energy, only. We can therefore express the heating rate in terms of a function of one variable that achieves its maximum when the total Alfv\'en wave energy is partitioned in a ratio of \( 2\!:\!1 \). By assigning wave energy according to this ratio, we can derive an efficiency ratio between Alfv\'en and kink wave heating to be $2\sqrt{3}/9 \approx 0.385$. This factor indicates that, even under conditions of optimal reflection, the Alfv\'en wave heating rate would be just over one-third of that of the kink wave, assuming equal total energy content. It should be noted that in practice the integrated Alfv\'en wave energy is not equal to the kink wave energy, as reflection tends to retain Alfv\'en wave energy within the domain for a larger amount of time. Taking this into account, these calculations provide Alfv\'en waves with every opportunity to outperform kink waves.

\begin{figure}[htp]
   \centering
    \includegraphics[width=.49\textwidth]{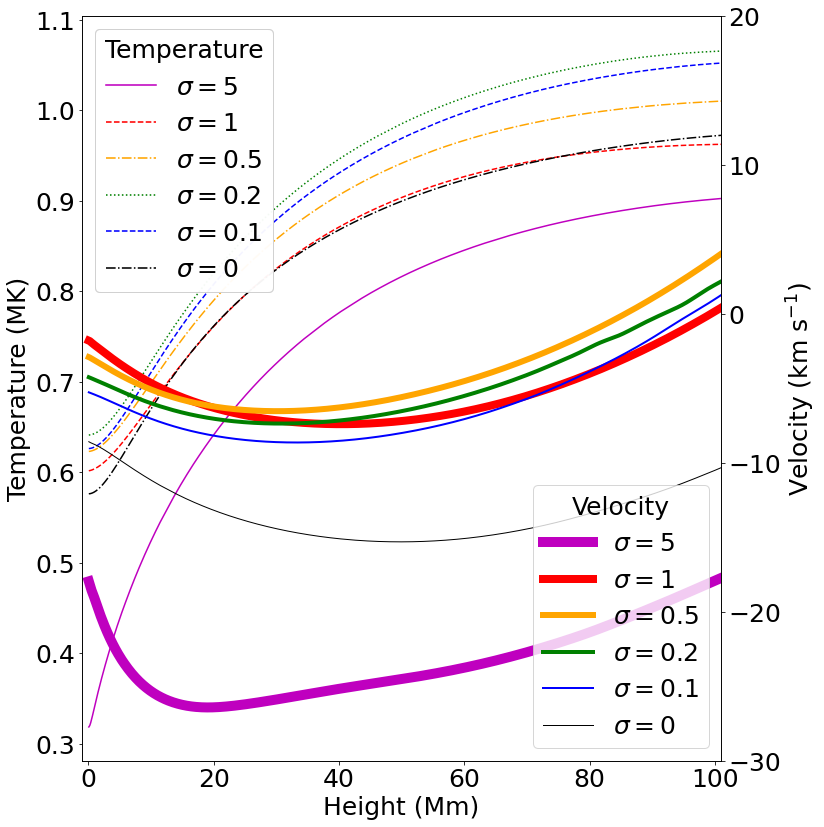}
    \caption{The same set up as Figure \ref{fig:AMRVAC_varyAWK_ref_Tv1_sameinj} but with increased Alfv\'en wave energy by the factor $9/2\sqrt{3}$.}
  \label{fig:AMRVAC_varyAWK_ref_Tv1}  
\end{figure}

\begin{figure}[htp]
   \centering
    \includegraphics[width=.45\textwidth]{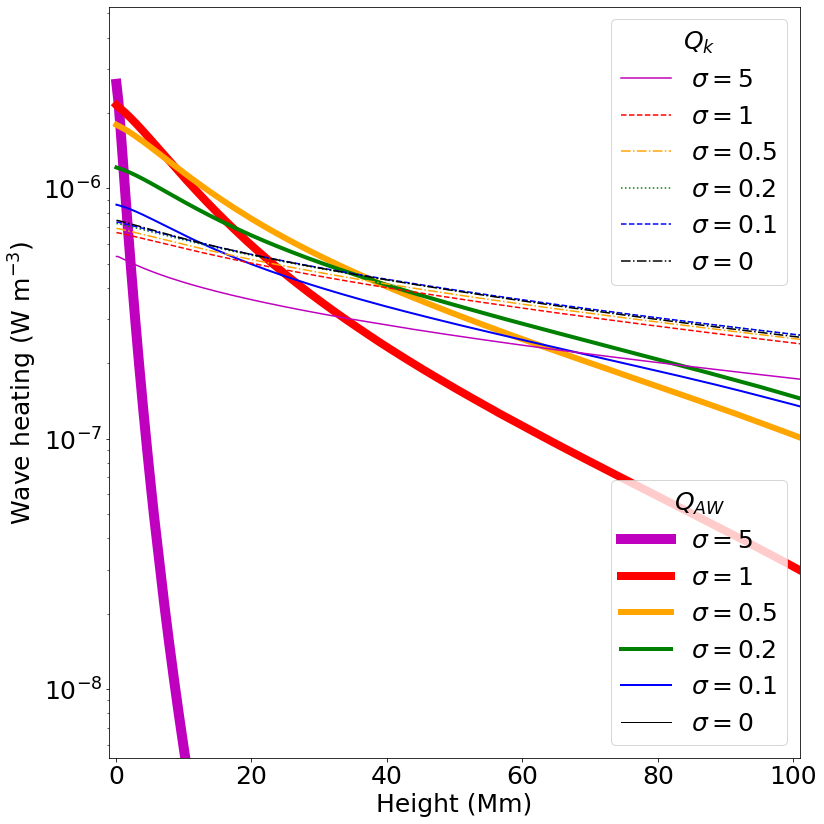}
    \caption{The same setup as Figure \ref{fig:AMRVAC_varyAWK_ref_heating_sameinj} but with increased Alfv\'en wave energy by the factor $9/2\sqrt{3}$.}
  \label{fig:AMRVAC_varyAWK_ref_heating}  
\end{figure}

We now conduct similar simulations to Section \ref{sec:sameAWK} whereby we keep the energy injections the same between simulations (with the increased Alfv\'en wave energy injection ), and vary our reflection coefficient, $\sigma$. Figures \ref{fig:AMRVAC_varyAWK_ref_Tv1} and \ref{fig:AMRVAC_varyAWK_ref_heating} present the temperature and velocity profiles and the heating components of Alfv\'en waves and kink waves as a function of height, respectively, at $t = 1800$ s. 

We find that there is an optimum reflection coefficient to generate the highest temperature plasma that relates to the green line ($\sigma = 0.2$) in Figure \ref{fig:AMRVAC_varyAWK_ref_Tv1}. In addition to this, we once again find that extreme levels of reflection (for $\sigma = 5$, represented by the pink line) cause an unstable influence on the plasma. Here, the Alfv\'en wave energy is concentrated around the base of the domain and disrupts the stability of the atmosphere, resulting in catastrophic cooling of the plasma. As discussed, Alfv\'en waves have a lower efficiency compared to kink waves in terms of their ability to heat a plasma, and now it has been shown that a corresponding increase in their energy injection cannot account for this inefficiency.

\section{Parameter study: varied Alfv\'en to kink energy}\label{sec:5}

In this section, we fix the reflection rate, $\sigma$, for Alfv\'en waves and vary the injected Alfv\'en and kink wave energy ratio at the base of the domain, such that the total energy injection at the base of the domain is equal throughout all simulations. One could argue that due to reflection, the cases where more Alfv\'en wave energy is considered, energy remains in the system for longer before propagating out of the open boundaries, allowing the Alfv\'en wave energy more time to heat the plasma. However, until an accurate reflection rate for kink waves is obtained, we cannot mitigate this effect but simply bear it in mind. We settled on a value of $\sigma = 0.5$ for the reflection coefficient of Alfv\'en waves since this value is consistent with the aforementioned literature. With that being said, we now conduct a parameter study whereby we vary the relative energy injection of the Alfv\'en and kink waves present in our simulation. 

\begin{figure}[htp]
   \centering
    \includegraphics[width=.45\textwidth]{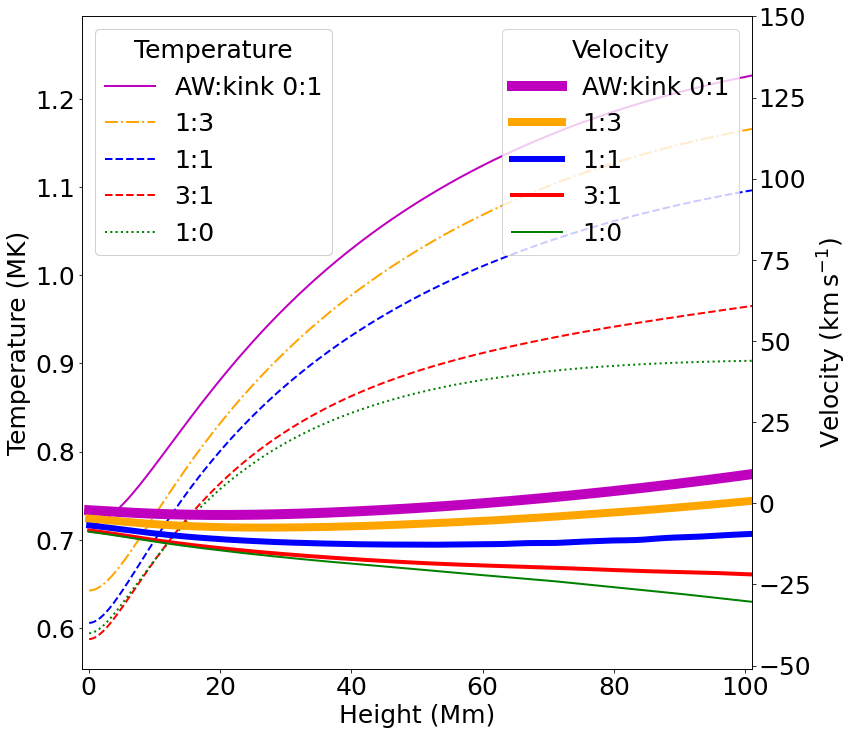}
    \caption{The temperature and velocity are shown as functions of height (Mm) plotted at time $t = 1800$ s, where the ratio of Alfv\'en to kink wave energy is varied. The various ratios (Alfv\'en : kink) are 0:1 (red), 1:3 (orange), 1:1 (green), 3:1 (blue) and 1:0 (black). The various thicknesses of lines refer to the velocity, and the various line styles refer to the temperature.}
  \label{fig:AMRVAC_varyAWK_ratio_Tv1}  
\end{figure}

\begin{figure}[htp]
   \centering
    \includegraphics[width=.45\textwidth]{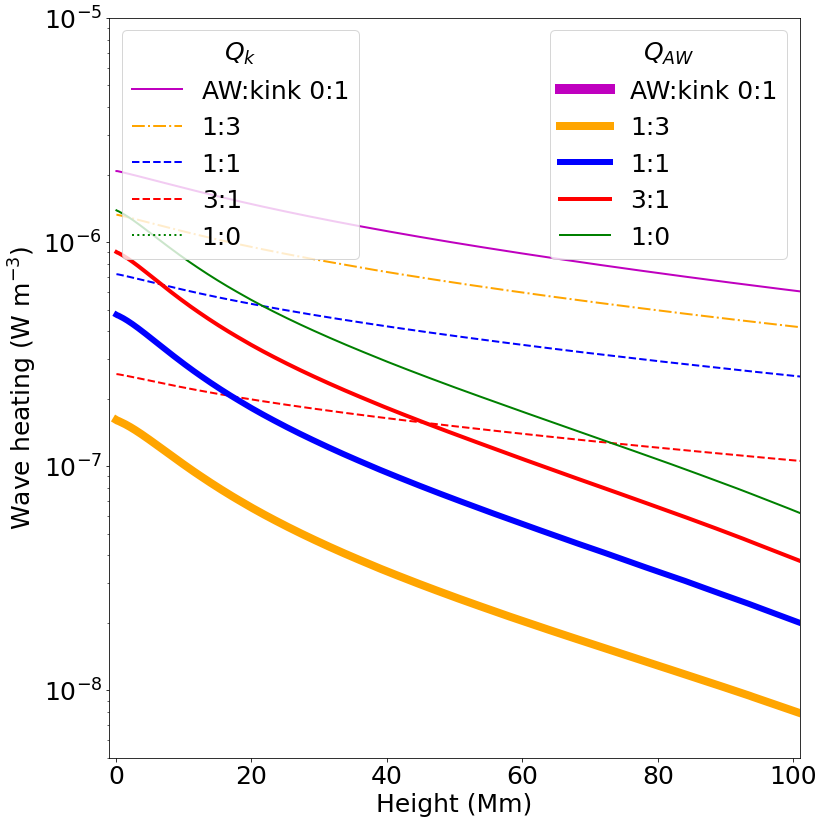}
    \caption{The heating rates corresponding to Alfv\'en and kink waves shown as functions of height for the various Alfv\'en to kink wave energy ratios. The various ratios (Alfv\'en : kink) are 0:1 (red), 1:3 (orange), 1:1 (green), 3:1 (blue) and 1:0 (black). The various thicknesses of lines refer to the Alfv\'en wave heating term, and the various line styles refer to the kink wave heating terms.}
  \label{fig:AMRVAC_varyAWK_ratio_heating}  
\end{figure}

Each line in Figure \ref{fig:AMRVAC_varyAWK_ratio_Tv1} represents a different ratio of Alfv\'en to kink wave energy. Red represents a simulation with only kink wave heating, orange represents a simulation with identical total energy but now spread across Alfv\'en and kink wave energy such that there is three times as much kink wave energy as Alfv\'en wave energy (3:1 ratio), the green line represents a balanced input of Alfv\'en and kink wave energy (1:1), the blue line represents a simulation where the Alfv\'en wave energy is dominant over kink wave energy (1:3 ratio) and lastly, the black like represents a simulation with only Alfv\'en wave energy. Below, Figures \ref{fig:AMRVAC_varyAWK_ratio_Tv1} and \ref{fig:AMRVAC_varyAWK_ratio_heating} show the temperature and velocity and the relative heating rates for each wave component as a function of height, respectively.

Previously, we mentioned that Alfv\'en waves alone cannot maintain an open field coronal plasma without the inclusion of an anomalous background heating function. This is why, in Figures \ref{fig:AMRVAC_varyAWK_ratio_Tv1} and \ref{fig:AMRVAC_varyAWK_ratio_heating}, we consider times before catastrophic cooling has had time to take effect. The Alfv\'en waves alone do not heat the atmosphere sufficiently, resulting in condensations (within the domain) with net inflow (negative) velocities. This considerable inflow causes the mass to drain out of the upper atmosphere, causing a collection of mass at the bottom boundary that cools catastrophically because the wave heating cannot compensate for the increased radiative losses. 

We would like to draw the reader’s attention back to the discussion on the radiative cooling timescale in our simulations. Throughout Sections \ref{sec:4} and \ref{sec:5}, we have shown heating rates that fall below the estimated \(10^{-5}\,\mathrm{W\,m^{-3}} \) required to balance the radiative losses of the solar corona. This explains why the lower regions of our atmosphere fail to sustain temperatures around 1 MK. However, near the upper boundary—where the density drops significantly—we find the radiative losses decrease to approximately \(10^{-7}\,\mathrm{W\,m^{-3}} \). The kink wave heating rates in these regions exceed these lower losses, which helps maintain stable heating and allows the solar atmosphere to reach equilibrium.

\section{Open field evolution: 1-3 $R_{\odot}$}\label{sec:LargeDom1}

Having demonstrated that wave heating can efficiently sustain an atmosphere extending up to $100$ Mm, it is natural to ask whether the same mechanism can operate on even larger scales. In particular, the solar corona itself extends far beyond $100$ Mm, and understanding the mechanisms that maintain its high temperatures out to distances comparable to the solar radius remains a central challenge in solar physics. In this section, we extend our model to investigate wave heating in an atmosphere reaching 3 $R_{\odot}$. Due to the extended atmosphere, we have to reduce the initial wave energy profile presented in Figure \ref{fig:Wkm_IC} at large altitudes in the domain through the use of an exponential function. In addition to this, our domain now greatly exceeds the predicted length scale of gravity in the solar atmosphere and hence, we now consider it to also vary with height. This assumption alters our boundary conditions, in that now gravity varies between the upper and lower boundaries, and the remaining changes are given below:

\begin{align}
    \vec{g}(z) &= -274 \text{m}\,\text{s}^{-2}\frac{R_\sun^2}{\left(z + R_\sun\right)^2}\ \vec{1}_z, \\
    W^-_\mathrm{k} &=
    \begin{cases}
        0.4 W_\mathrm{k0}\left(\cos\left(\dfrac{\pi z}{\delta}\right) + 1.5\right), & \text{for } z < \delta, \\
        0.2 W_\mathrm{k0} \exp\left(\dfrac{-(z-\delta)}{20\,\mathrm{Mm}}\right), & \text{otherwise},
    \end{cases} \\
    W^+_\mathrm{k} &= W^\pm_\mathrm{A} = 0,
    \label{eq:large_dom_setup}
\end{align}
where $W_\mathrm{k0} = 5.3\times10^{-3}\,\text{J}\,\text{m}^{-3}$ and the scale height of $20$ Mm for the exponential. The choice of profile for the initial wave energy in the domain is arbitrary, but allows the code to run stably.

\begin{figure}[htp]
   \centering
    \includegraphics[width=.49\textwidth]{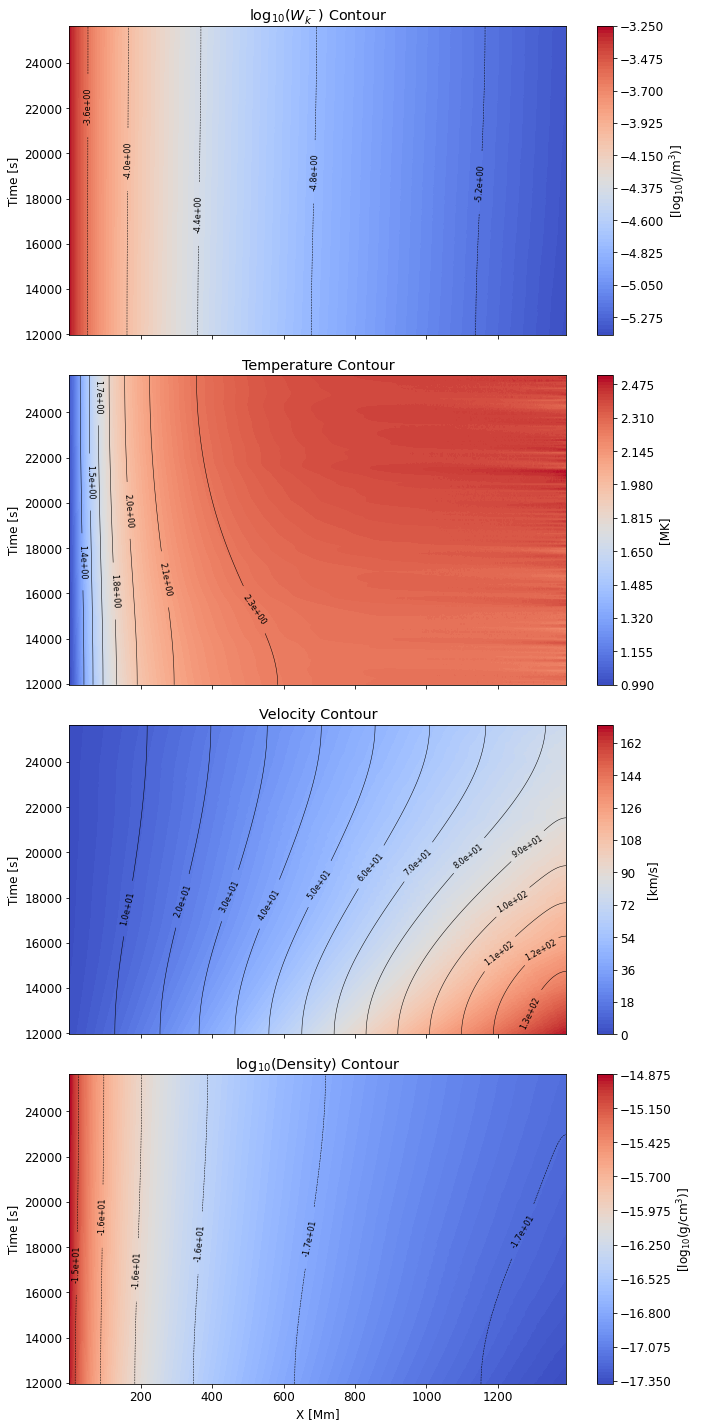}
    \caption{\texttt{MPI-AMRVAC} simulation as prescribed in Equations (\ref{eq:large_dom_setup}). We consider this a qausi-steady state. The quantities shown are the outward propagating kink wave energy (scaled by $\log_{10}$, in units $\mathrm{J\,m^{-3}}$), temperature (in units MK), the velocity (in km\,s$^{-1}$) and the plasma density (scaled by $\log_{10}$, in units g\,cm$^{-3}$) as functions of time (seconds) and space, $X$ (Mm).}
  \label{fig:Kink_only_2RSun_Gstrat}  
\end{figure}

In Figure \ref{fig:Kink_only_2RSun_Gstrat}, we choose to showcase the stable atmosphere maintained by kink waves alone, since this setup provided the most stable atmosphere, and showcases best the efficiency of kink wave heating through uniturbulence.

In our simulations, it is necessary to evolve the system for a sufficient duration to ensure that artifacts arising from the initial conditions are advected out of the computational domain, often in the form of a shock wave. To avoid confusion with these transient events, we present our results after this shock has propagated beyond the domain boundaries, focusing instead on the (quasi-)steady state that subsequently develops. Notably, this initial shock does not arise in simulations with the smaller domain size, which we attribute to the lower densities and pressures characteristic of the larger domain setup. Additionally, during the advection of the shock wave, the larger domain exhibits a net mass outflow — a feature far less pronounced in the smaller domain — which leads to a modified steady-state compared to the smaller domain. Since the rate of wave energy injection is constant across all simulations, the differences in mass retained in each domain can lead to differences in temperatures and outflow velocities. Thermal conduction redistributes this heating throughout the domain, leading to global differences in the temperature profile. That being said, comparing Figures \ref{fig:Kink_only_2RSun_Gstrat} and \ref{fig:AMRVAC_kink_only_steady}, the temperature at the top boundary matches very well with the temperature at 100 Mm in the much larger 1-3 $R_{\odot}$ simulation, suggesting these discrepancies have only a minor effect on the overall solution.

\section{Notes on the UAWSoM model}\label{sec:Notes}

In this investigation, we have made some general simplifications To make progress and improve upon the widely used AWSoM model. In this section, we address the potential impacts these assumptions may have on the model.

\subsection{Reflection rates for kink waves}

We omit the study of the reflection of kink waves. There has been, to the best of our knowledge, no investigation able to quantify the reflection a kink wave undergoes due to atmospheric stratification. With that being said, kink waves do not require counterpropagating waves to heat the plasma, as they undergo self-cascade via uniturbulence. Therefore, in the current setup, reflection of kink waves will only work to redistribute the location of heating, rather than guarantee it, as is the case for Alfv\'en waves. If kink wave reflection were to be modeled analogously to Alfv\'en wave energy reflection—such that it were proportional to the longitudinal gradient in wave speed—it would likely concentrate heating lower in the atmosphere, where gradients in the magnetic field and density are steepest. This, in turn, would most likely yield a temperature profile that is closer to being isothermal than the steady state shown in Figures \ref{fig:AMRVAC_kink_only_steady} and \ref{fig:Kink_only_2RSun_Gstrat} as has been observed by \cite{Aschwanden2000}, who found EUV loops that consist of near-isothermal threads with a substantially smaller temperature gradient than is predicted by the Rosner-Tucker-Vaiana scaling law model with gravitational stratification incorporated \citep[][known as the RTVS scaling law]{Rosner1978, Serio1981}.

In addition to kink wave reflection, it is also likely that Alfv\'en and kink waves will interact both with themselves (i.e., in the case of self-interaction of Alfv\'en waves, we take this into account already) as well as one another nonlinearly, resulting in the expedition of the dissipation of wave energy. This has been shown by \cite{Guo2019} whereby Alfv\'en and kink waves, when in the presence of one another, expedited the onset of the Kelvin-Helmholtz instability. In our simulations, this could be quantified by introducing a new correlation length for kink waves that have coupled to Alfv\'en waves, e.g., $L_{\perp,\mathrm{k,AW}}$ which may multiply term(s) proportional to some power and combination of $W_\mathrm{k}^\pm$ and $W_\mathrm{A}^\mp$. It is also likely that the relative power in the Alfv\'en and kink waves would play a role in this dissipation length. This is also beyond the scope of the present study, however, we aim to address these coupling effects in future work. In the present investigation, we instead focus on the separate roles that kink waves and Alfv\'en waves have on the atmosphere.

\subsection{Resonant absorption}

In our model, we have considered a density inhomogeneity represented by a discontinuity with a given magnitude, $\zeta$. Discontinuities are a mathematical simplification and clearly do not exist in the solar atmosphere, and should be replaced by a density gradient. In this case, there exists a gradient in the Alfv\'en speed, which can lead to resonances, resulting in a transfer of energy from the global kink mode into torsional Alfv\'en waves in the resonant layer. These Alfv\'en waves are subject to phase mixing and/or turbulent dissipation. To quantify the distribution of kink wave energy transferred to Alfv\'en waves and the portion available for uniturbulent dissipation, it is useful to calculate the time scales of these two processes. \cite{TVD2020} calculated the time scale of uniturbulence ($\tau_{\mathrm{U}}$) to be 

\begin{equation}
    \tau_{\mathrm{U}} = \sqrt{5\pi}\frac{R}{V}\frac{2(\zeta+1)}{\zeta-1},
\end{equation}
where $R$ is the radius of the plume (or coronal loop in their case), $V$ is the velocity amplitude of the oscillation. The time scale of resonant absorption ($\tau_{\mathrm{RA}}$) is given by, e.g., \cite{Goossens1992, Ruderman2002} as

\begin{equation}
    \tau_{\mathrm{RA}} = \frac{1}{\pi}\frac{R}{l}\frac{2(\zeta+1)}{\zeta-1}P,
\end{equation}
where $l$ is the width of the resonant layer and $P$ is the period of the oscillation. Note that the formulae have been adapted to match the notation consistent with the present investigation. 

Given the plasma parameters employed throughout this investigation, we find that $\tau_{\mathrm{U}} \approx 12000/V$ s, and $\tau_{\mathrm{RA}} \approx 1000P/l$ s. Given a velocity amplitude of $10$ km\,s$^{-1}$, a period of $400$ s \citep{TVD2020} and resonant layer $100$ km in width \citep[satisfying the condition $l/R \ll 1$,][]{Goossens1992}, the two time scales are given by $\tau_{\mathrm{U}} \approx 1200$ s, and $\tau_{\mathrm{RA}} \approx 3800$ s. This suggests that uniturbulence is a faster acting mechanism than resonant absorption, with only a relatively small amount of kink wave energy being deposited in the resonant layer in the form of Alfv\'en waves before uniturbulence dissipates the kink wave energy. We should preface that these calculations are done assuming a coronal loop geometry, rather than open field lines, however, we propose that uniturbulence would remain the dominant dissipation mechanism. In future work, we intend to parameterise this transfer of energy from the kink wave to the Alfv\'en wave to capture this effect.

\section{Conclusions and future Work}\label{sec:Conclusions}

In this study, we have extended previous wave-based coronal heating models by implementing a new physics module into the \texttt{MPI-AMRVAC} framework—UAWSoM—that includes both Alfv\'en and kink wave dynamics within an MHD framework. Unlike traditional models driven by Alfv\'en waves alone, which require an ad hoc background heating term to maintain coronal stability, our kink wave-driven simulations demonstrate that a stable 1D model atmosphere can be sustained purely through kink wave dissipation. This highlights the inherent efficiency of uniturbulent dissipation as a heating mechanism, particularly in the lower solar atmosphere where transverse density gradients persist, enhancing the damping of kink waves.

Furthermore, we find that Alfv\'en waves alone are insufficient to maintain stability in our model, irrespective of the magnitude of the reflected component. In fact, under conditions of strong reflection, Alfv\'en waves introduce a destabilising influence, resulting in the onset of catastrophic cooling. In contrast, kink waves provide a more robust source of heating, capable of operating in the absence of counterpropagating wave components. Our results suggest that a hybrid model involving both Alfv\'en and kink waves can be viable, but only if sufficient kink wave energy is present to compensate for the limitations of Alfv\'en wave-based heating. We propose that it is likely that kink wave heating is essential in a self-consistent AC heated corona. These findings offer a potential physical explanation for the ad hoc background heating functions often employed in AWSoM-type models \citep{Reville2020}. 

We now wish to acknowledge some limitations of the modelling performed in this investigation. The base magnetic field and density contrast that were chosen are larger than estimated in the quiet Sun and coronal holes \citep[e.g., see][]{Morton2021}, but allow for a stable atmosphere to be found. In addition, we do not include additional cooling contributions such as strong thermal conduction down to a cooler transition region and chromosphere, which would lead to enhanced cooling at the footpoint from transition region material. We intend to include these additional effects in future works. In addition to these limitations, the inclusion of more complex wave modes is essential in coronal heating and solar wind driving. Future work will include the exploration of additional wave modes, such as the slow-mode wave and their coupling to parametric instabilities \citep[see, e.g.,][]{Shoda2018, Shoda2019}, further enhancing the realism and completeness of wave-based solar atmospheric models. In addition, as shown by \cite{Guo2019}, kink waves and Alfv\'en waves do interact nonlinearly, leading to an earlier onset of the Kelvin-Helmholtz instability, and this is likely to play a role in reducing the correlation lengths of both wave modes. 

Building on our current advancements, we propose to extend our modeling efforts to global 3D simulations of the solar atmosphere, initiated from observed magnetograms. Using field-line extrapolation techniques, such as the Potential Field Source Surface (PFSS) model, we will reconstruct the large-scale magnetic structure extending from the photosphere into the corona. This approach will allow us to naturally incorporate both open field regions (e.g., coronal holes) and closed magnetic loops (e.g., active regions and quiet Sun structures) into our simulations.

We aim to systematically investigate the heating of both active regions and quiet Sun areas, testing whether the turbulent cascade of kink wave energy can account for the varied thermal structures observed across the solar corona. To bridge simulation results with observations, we also plan to employ forward modeling techniques using tools such as FoMo \citep{TVD2016} to synthesise observable quantities (e.g., intensity maps, Doppler shifts, line broadening) from our simulation outputs. This will enable direct comparison with data from instruments like SDO/AIA, Hinode/EIS, and Solar Orbiter/SPICE. We expect these global simulations will provide a powerful test of whether kink-wave-turbulence, when combined with realistic magnetic topologies, can consistently heat the solar atmosphere and drive the solar wind across a range of solar conditions.

We can use the same approach as for the Sun to model stellar atmospheres and winds of low-mass stars, as they share similar characteristics. A recent study by \cite{Vidotto2023} proves the importance of using global magnetically-driven 3D models such as AWSoM to model the stellar wind, in this case around a cool dwarf, and its interaction with the orbiting exoplanet, although uncertainties in input parameters and the corresponding output, such as predicted mass-loss rate, are still large. We believe that the UAWSoM model presented here can shed more light on the stellar wind characteristics and star-planet interactions.

\begin{acknowledgements}
MM, TVD and LB received financial support from the Flemish Government under the long-term structural Methusalem funding program, project SOUL: Stellar evolution in full glory, grant METH/24/012 at KU Leuven. Furthermore, TVD and DL were supported by a Senior Research Project (G088021N) of the FWO Vlaanderen. The research that led to these results was subsidised by the Belgian Federal Science Policy Office through the contract B2/223/P1/CLOSE-UP. It is also part of the DynaSun project and has thus received funding under the Horizon Europe programme of the European Union under grant agreement (no. 101131534). Views and opinions expressed are, however, those of the author(s) only and do not necessarily reflect those of the European Union and therefore the European Union cannot be held responsible for them.
\end{acknowledgements}

\bibliographystyle{aa}

\bibliography{ref.bib}.

\end{document}